\documentstyle[aps,prl,multicol]{revtex}
\begin{document}

\draft

\title{Base pair dynamic assisted charge transport in DNA}

\author{E. I. Kats $^{1,2}$ and V. V. Lebedev $^{1,3}$}

\address{$^1$ L. D. Landau Institute
for Theoretical Physics, RAS \\
117940, Kosygina 2, Moscow, Russia; \\
$^2$ Institut Laue-Langevin, 6 rue Jules Horowitz,
BP 156, Grenoble, France; \\
$^3$ Department of Physics, Weizmann Institute of Science,
76100, Rehovot, Israel.}

\date{\today}

\maketitle

\begin{abstract}

An $1d$ model with time-dependent random hopping is proposed to
describe charge transport in DNA. It admits to investigate both
diffusion of electrons and their tunneling between different sites
in DNA. The tunneling appears to be strongly temperature-dependent.
Observations of a strong (exponential) as well as a weak distance
dependence of the charge transfer in DNA can be explained in the
framework of our model.

\end{abstract}


\begin{multicols}{2}

Electronic transport is a ground to a wide range of important
biological processes in DNA. Besides, the phenomenon has a
fundamental physical interest, since the transport properties of
biomolecules are expected to differ considerably from those of
macroscopic conductors. And at last, very recently material
scientists also turned their attention to charge migration in DNA
for the development of DNA-based molecular technologies.

Although first attempts to measure DNA conductivity \cite{ES62}, and
to give a theory of the phenomenon \cite{lad0} have been made almost
40 years ago, the question concerning charge transport through DNA
remains unsettled, and there is an impressive quantity of unexplained
or partially explained data. Different publications report frequently
contradictory results. Two kinds of techniques for getting information
on charge transport in DNA are used. First, direct or indirect
electrical conductivity measurements on micrometer-long DNA ropes are
performed \cite{j1,super,j2,j?,j3}. Experimental results obtained in
this technique are ambiguous. DNA conductivity $\sigma$ was reported
as almost metallic of the order $10^4\,\Omega^{-1}cm^{-1}$ \cite{j1}
(in a recent publication \cite{super} the authors claim they observed
even proximity induced superconductivity in DNA) or semiconducting
with $\sigma\simeq0.1\,\Omega^{-1}cm^{-1}$ \cite{j?}. Very recently
experimental techniques have progressed to the point where the
conductivity on individual $10nm$ long double stranded molecules was
measured \cite{j3} and the result implies that DNA is a good
insulator. Clearly this frustrating situation with conductivity
measurements means that there are many relevant factors which can
influence the charge transport in DNA by different way and which are
hardly controlled in real experiments. The second technique, related
to fluorescence quenching measurements on DNA strands, doped with
donor and acceptor molecules
\cite{b1,b2,b3,b4,b5,b6,b7,b8,b9,b10,b11,b12,FP95,FT98,b13,GA01},
seems more reliable, and it is our main concern here. In this
technique photo-excitation of a donor associated with the stack of
base pairs in some fashion, allows transfer of an electron to the
stack. The migrating electron is trapped finally at the acceptor site,
and charge transfer is monitoring by the yield of a chemical reaction
accompanying the trapping process. The transfer rate is usually
assumed to be characterized by a simple exponential law $\exp(-\beta
x)$, where $x$ is the donor-acceptor separation. Fitting to this law
gives values of $\beta$ ranging from $0.1\AA^{-1}$ to $1.4\AA^{-1}$.

It is a common wisdom that DNA can be treated as an one-dimensional
linear chain of stacked base pairs. We believe, that in the ground
state every base pair contains bound electrons only. Then the charge
is carried through DNA by excited electrons (or holes) which can jump
between the base pairs. Below we propose a simple model which, to our
meaning, reflects basic features of the electron transport in DNA. The
picture includes the following ingredients:

\noindent
(i) The excited electron states at the base pairs are separated by
energy spacing larger than temperature and therefore thermally are
practically not excited;

\noindent
(ii) Thermal motions of the DNA base pairs are elastic vibrations
with a characteristic frequency $\omega_b$;

\noindent
(iii) Efficient charge transfer between neighboring base pairs takes
place for rare events;

\noindent
(iv) The Coulomb interactions between the electrons and holes can be
neglected for the description of hopping transport.

\noindent
Let us explain the point (iii) in more detail. For the static
equilibrium DNA helix charge hopping is expected to be negligibly
small since there is no significant electronic overlapping between
adjacent base pairs. Nevertheless, sometimes, due to thermal
fluctuations, exclusively favourable for hopping configurations of
the base pairs occur, when an efficient hopping is possible. If the
separation between the pairs is larger than the amplitude of their
thermal vibrations, then probability of such events (which can be
called ``contacts") is small. Duration of the contact can be
estimated as the characteristic oscillation time $\omega_b^{-1}$.

Now we discuss a correspondence of the assumptions enlisted above and
experimental data. As a guide line we use not only data known from the
DNA literature but as well the data obtained for a wide range of
organic linear chain polymers of stacked planar molecules (for a
review see \cite{JS82}). The contacts are related to mutual
displacements and orientations of adjacent base pairs. Probably,
hopping matrix elements are mostly sensitive to the relative rotations
(twist fluctuations) of the base pairs (see, e.g., \cite{BG00}). The
characteristic frequency of these fluctuations, $\omega_b$, is usually
estimated as being in the region $10^{11}\div10^{12}\,s^{-1}$. A small
probability of the contacts is confirmed by experiment showing that
the characteristic electronic hopping time $\tau$ is larger than
$\omega_b^{-1}$, in the experiments \cite{b1,b3,b10,b11}
$\omega_b\tau= 10^2\div10^3$.  Our first assumption (i) requires
$\Delta E>T$, where $\Delta E$ is the spacing in the spectrum of
electron excitations for a base pair. The magnitude of $\Delta E$ can
be measured directly, for the experiments
\cite{b1,b2,b3,b4,b5,b6,b7,b8,b9,b10,b11,b12,FP95,FT98,b13,GA01} 
$\Delta E> 500 K$. Therefore the inequality is satisfied. Rough
macroscopical estimations of the Coulomb interaction $U_c$, as well as
ab initio molecular orbital calculations of $U_c$, give few $meV$
\cite{DF00}, and thus Coulomb energy appears to be smaller (though of
the same order) than $\hbar\omega_b$. We believe that it is enough to
justify neglecting Coulomb interaction.

The above reasoning leads to an $1d$ hopping Hamiltonian for the
electrons
\begin{eqnarray} &&
{\cal H}=\hbar\sum_i \left(\xi_i a_i^+a_{i+1}
+\xi_i^* a_{i+1}^+a_{i}\right) \,.
\label{dnb1} \end{eqnarray}
Here $a_i$ and $a_i^+$ are electronic annihilation and creation
operators at the site (i.e. the base pair) with the number $i$, 
and $\xi_i$ are the hopping amplitudes, which are time-dependent
quantities. The equations for the Heisenberg operators $a_i$ are
\begin{eqnarray} &&
\partial_t a_i=-i\xi_i a_{i+1}
-i\xi_{i-1}^* a_{i-1} \,.
\label{dnbb} \end{eqnarray}
We assume that different $\xi_i$ possess independent statistics,
since $\xi_i$ are related to independent thermal pair base
fluctuations. The hopping matrix element $\xi_i$ can be decomposed
into a constant part $\langle\xi_i\rangle$, that describes the
coherent charge carrier motion in a completely rigid lattice, and a
fluctuating part. Since the probability to jump is appreciable
during rare events, the coherent part of $\xi$ can be neglected in
comparison with its fluctuating part.

Note, that theoretical models based on hopping Hamiltonians similar
to Eq. (\ref{dnb1}) are widely used to describe charge transport in
solid state physics (see, e.g., \cite{zim,appel}). For most of
problems in this case the description corresponds to electron
migration in a steady energy landscape, including thermally
activated jumps over barriers and quantum tunneling through the
barriers. It is quite different from our case.

We assume that DNA molecules can be treated as homogeneous ones.
Though the molecules are constructed from four different nitrous
bases, experimental data
\cite{b1,b2,b3,b4,b5,b6,b7,b8,b9,b10,b11,b12,FP95,FT98,b13}, as well
as numerical first principle calculations \cite{DF00,j4,lad} show,
that the sequence of base pairs is not a decisive factor which
determines electron transport in DNA. Quantitatively this condition
can be formulated as $\delta E<\hbar\omega_b$, where $\delta E$ is an
energy spacing between the (lowest excited) electron energy levels at
different base pairs. The values of $\delta E$, known mainly from
numerical electronic structure calculations \cite{DF00,j4,lad}, are of
order of $meV$. Thus $\delta E$ is smaller than $\hbar\omega_b$ for
$\omega_b$ given above, that justifies the picture. Besides some
experiments (see, e.g., \cite{GA01}) are performed for artificial
homogeneous DNA, where $\delta E=0$. One expects that, due to the
hopping, electron diffusion occurs on large time scales. For $\xi$,
treated as a white noise, it was demonstrated in the papers
\cite{HR72,HS73}. Though our case is essentially different, there is
good reason to believe, that the same behavior should be observed on
time scales larger than the hopping time $\tau$.

Below, we examine the particular case related to the fluorescence
measurements, reported in the papers
\cite{b1,b2,b3,b4,b5,b6,b7,b8,b9,b10,b11,b12,FP95,FT98,b13,GA01}. The
donors are photo-excited and effects, related to the excited electron
motion to the acceptors, are monitored. The energetic gaps $\delta
E_d$ and $\delta E_a$ between the donor and the acceptor and the base
pairs between them, are crucial for the hopping rate.  The values of
$\delta E_d$ and $\delta E_a$ (known mainly from ab initio numerical
calculations \cite{b8,DF00,j4,lad}) can be estimated as $10^2 meV$. We
see, that the inequalities $\delta E_d,\delta E_a\gg\hbar\omega_b$ are
satisfied. The electron is always bound to the acceptor site more
strongly than to a standard base pair, that is $\delta E_a>0$. As to
the donors, the sign of $\delta E_d$ can be either positive or
negative. If $\delta E_d$ is negative then a scheme of the electronic
charge transfer from the donor to the acceptor is quite simple.
Initially, the electron leaves the donor, jumping to the neighbor
site, and then jumps between the standard base pairs, trapping finally
at the acceptor. The case $\delta E_d>0$ is more complicated. In order
to have a driving force for the donor--acceptor charge transfer
process the final state with the charge bound to the acceptor should
be energetically favourable, that is the inequality $\delta E_a>\delta
E_d$ has to be satisfied. However, there are some base pairs
in-between which play the role of the potential barrier for the
electron. Therefore there are two possibilities for the electron to
come to the acceptor. The first possibility is to jump initially from
the donor to the neighbor site and then to move to the acceptor due to
multistep hopping over the standard base pairs. The second possibility
is the unistep (direct) quantum tunneling from the donor to the
acceptor through the barrier.

Since $\delta E_d\gg\hbar\omega_b$, the probability for the electron
to jump from the donor to the neighboring base pair due to dynamics of
$\xi$ is negligible. At $\delta E_d>0$ such a jump is possible if the
electron absorbs a high-frequency phonon with the frequency
$\omega_{ph}\sim\omega_d$ ($=\delta E_d/\hbar$). Correspondingly, at
$\delta E_d<0$ the electron jump from the donor is accompanied by
emitting high frequency phonons. Such dynamical vibrations with
periods as short as tens femtoseconds (i.e. phonons with
$\omega_{ph}\sim 10^{14}s^{-1}$) were reported in the literature
\cite{HL84,BM98,HP01}. Since $\hbar\omega_{ph}>T$, occupation numbers
of such phonons are small. Thus, for $\delta E_d>0$ the probability
for the electron to jump from the donor to the neighboring site
contains two small factors: the probability of the contact and the
probability to absorb the high frequency phonon. It corresponds to the
experimental situation where only a small fraction of the electrons
are transported from the donor to the acceptor.

When the electron leaves the donor, it starts to jump between the
donor and the acceptor. It can return to the donor or can come to the
acceptor. If $\delta E_d<0$ then the probability to return to the
donor is negligible. We assume that even at $\delta E_d>0$ the
probability of the electron to jump to the donor or to the acceptor is
smaller than the probability to jump to the standard base pair. There
are two reasons for the assumption. First, the donors and the
acceptors have chemical structures, different from the standard base
pairs, that hinders for the contacts. Second, the jump has to be
accompanied by the phonon emission, that diminishes its probability.
The same is valid for the acceptor. Thus before being finally trapped
at the acceptor site, the electron jumps many times back and forth
over the base pairs between the donor and the acceptor, ``smearing
out'' homogeneously over the all intermediate base pairs. Then the
relative probability for the electron to come to the acceptor is
determined by the ratio of the probabilities for the electron to jump
to the donor and to the acceptor from adjacent base pairs. This
relative probability appears to be independent of the separation $x$
between the donor and the acceptor. That explains why the rate of
charge transfer sometimes is almost insensitive to the relative
loading of donors and acceptors (see, e.g., \cite{p1,p2}). The above
picture implies, that the total donor-acceptor charge transfer time
should be larger than the electronic hopping time $\tau$, and it
conforms to experimental data (see, e.g., \cite{b1}).

Now we consider the quantum tunneling for the electron, strongly
attached to the donor, that is the case $\delta E_d>0$. Though the
potential barrier depends on time, at the condition $\delta
E_d\gg\hbar\omega_b$ the probability for the electron tunneling from
the donor to the acceptor can be calculated in the adiabatic
approximation.  To examine the tunneling process, one should consider
the quasistationary electron state bound at the donor. In the spirit
of our picture we assume $\xi\ll\omega_d$. Then the energy of the
bound state is close to $-\delta E_d$. Substituting $\partial_t$ by
$i\omega_d$ in Eq. (\ref{dnbb}), one obtains for the state
\begin{eqnarray} &&
a_i=-\frac{\xi_{i-1}^*}{\omega_d}a_{i-1} \ {\rm if}\
i>0\,, \quad
a_i=-\frac{\xi_{i}}{\omega_d}a_{i+1} \ {\rm if}\ i<0 \,,
\label{solu} \end{eqnarray}
where we used the condition $\xi\ll\omega_d$. Then the probability
for the electron to be at the site $n$ is determined by the average
\begin{eqnarray} &&
\langle a_n^+ a_n \rangle
= \omega_d^{-2n} \left\langle \left| \prod_{i=1}^{n}
{\xi_{i-1}}\right|^2 \right\rangle \,.
\label{prob} \end{eqnarray}
Quantum averaging and averaging over statistics of $\xi$ are performed
at deriving Eq. (\ref{prob}). Besides at the derivation we substituted
$\langle a_0^+a_0\rangle\approx1$ justified by $\langle a_n^+ a_n
\rangle \ll1$. Note that the probability (\ref{prob}) is determined by
the simultaneous statistics of $\xi$. Remind that different $\xi_j$
are assumed to be statistically independent. Therefore the average in
the right-hand side of Eq. (\ref{prob}) is a product of
$\langle|\xi_j|^2\rangle$. For the standard pairs the quantities can
be regarded as site independent ones. Therefore the probability of the
electron to be at the nearest to the acceptor site is proportional to
$\langle|\xi|^{2}\rangle^n/{\omega_d^{2n}}$ where $n$ is the number of
the standard pairs between the donor and the acceptor.

The jump of the electron from the bound state to the acceptor is
accompanied by the phonon emission. However, the only $x$-dependent
factor in the probability of the process is related to the average
charge occupation number of a site $n$ near the acceptor, established
above. Thus, we obtain for the probability the exponential law
$\exp(-\beta x)$ with
\begin{eqnarray} &&
\beta=a^{-1}\ln({\omega_d^2}/{\langle|\xi|^2\rangle}) \,,
\label{beta} \end{eqnarray}
Here $\xi$ is the hopping probability for the standard base pairs and
$a=3.4\AA$ is the distance between the base pairs in DNA. Note, that
$\omega_d$ depends on the donor type, whereas the average
$\langle|\xi|^2\rangle$ is mainly related to base pair vibrations.  It
follows from the above consideration, that the exponential law implies
the condition $\beta>a^{-1}$. This conclusion is in agreement with the
majority of published experimental data.  Reported in \cite{FT98} the
value $\beta=0.1\AA^{-1}$ (thus smaller than $a^{-1}$) is, probably,
related to an attempt to fit a complex behavior (including two
processes: diffusion and tunneling) by a simple exponential law.

Let us stress that the quantum tunneling analyzed above is not a
standard (static) tunneling described in textbooks. We have deal with
dynamic tunneling which can be effective only when due to fluctuations
of $\xi$ there occurs some kind of a ``bridge'' from the donor to the
acceptor. The exponential law, we found, is explained in fact by a
small probability to have such a bridge, which is realized, when
simultaneously many contacts between base pairs occur. In addition,
the probability of this kind of tunneling is strongly dependent on the
temperature via $\langle|\xi|^2\rangle$. It is natural to assume that
$\xi$ exponentially depends on the relative displacement $u$ of the
neighboring base pairs. Then (in the harmonic approximation)
$\ln\langle|\xi|^2\rangle$ contains the term, proportional to $\langle
u^2\rangle$, which is proportional to the temperature $T$. Thus we
arrive at the expression
\begin{eqnarray}
\beta a=c_1-c_2 T \,,
\label{temp} \end{eqnarray}
where $c_1$ and $c_2$ are temperature-independent factors. They can be
extracted from the paper \cite{FP95}: $c_1\approx4$,
$c_2\approx0.01K^{-1}$. The values are in agreement with rough
estimates $c_1\sim a/b$, $c_2\sim k_B/(M\omega_b^2 b^2)$, where $k_B$
is the Boltzmann constant, $M$ is the base pair mass, and $b$ is an
electronic penetration length. It can be estimated as
$b\sim\hbar/\sqrt{mE}\sim1\AA$, where $m$ is is the electron effective
mass and $E$ is its binding energy at the base pair.

To conclude, for the electron, strongly bound to the donor, we
established two different charge transfer mechanisms: diffusion and
tunneling. The diffusion leads to the charge transfer probability
independent of the donor-acceptor distance $x$. However, the
probability contains the small factor related to the electron jump
from the donor to a neighboring site. The tunneling leads to the
exponential dependence of the probability on $x$ (with the
temperature-dependent length $\beta^{-1}$). Therefore, it is not
efficient for large distances. Thus, the exponential law has to be
observed for small distances $x$ whereas for large distances the
charge transfer rate has to be independent of the donor-acceptor
distance. Just this kind of behaviour was reported very recently
\cite{GA01}. For the case of the electron weakly bound to the donor
the hopping should always dominate over quantum tunneling. That
explains why the rate of charge transfer sometimes does not behave
exponentially even for small $x$ \cite{p1,p2}.

Note that in some cases the interaction of light, ionizing radiation
or chemically active reagents with DNA can result in loss of an
electron at a specific site with formation of a hole. In this case the
charge transport through DNA can be provided by holes (see, e.g.,
\cite{b7}). The key issues for positive charge carrier transport are
the same as for the electrons. As far as the physical picture of
charge transport is essentially the same for both kind of carriers, it
can be described in the framework of the same approach.

The research presented in this publication was made possible in
part by RFFR Grant 00-02-17785. Fruitful discussions with A.
Iosselevitch, Yu. Evdokimov, V. Golo, D. Bicout, and T. Costi are
gratefully acknowledged.

\end{multicols}

\end{document}